\documentclass[pdflatex,sn-mathphys-num]{sn-jnl-my}
\usepackage{graphicx}%
\usepackage{multirow}%
\usepackage{amsmath,amssymb,amsfonts}%
\usepackage{amsthm}%
\usepackage{mathrsfs}%
\usepackage[title]{appendix}%
\usepackage{xcolor,soul}%
\usepackage{textcomp}%
\usepackage{manyfoot}%
\usepackage{booktabs}%
\usepackage{algorithm}%
\usepackage{algorithmicx}%
\usepackage{algpseudocode}%
\usepackage{listings}%
\usepackage{listings}
\usepackage{color}
\usepackage{placeins}
\usepackage[justification=centering]{caption}

\definecolor{dkgreen}{rgb}{0,0.6,0}
\definecolor{gray}{rgb}{0.5,0.5,0.5}
\definecolor{mauve}{rgb}{0.58,0,0.82}

\theoremstyle{thmstyleone}%
\theoremstyle{thmstyletwo}%

\theoremstyle{thmstylethree}%

\begin{document}

\title[]{Deriving thin-film averaged equations using computer algebra
% : the two-phase WRIBL model and SymPy
}


\author[1]{\fnm{Swarnaditya } \sur{Hazra}}\email{swarnaditya.hazra@gmail.com}

\author*[1]{\fnm{Jason R.}\sur{Picardo}}\email{picardo@iitb.ac.in}
% \equalcont{}

\affil[1]{\orgdiv{Department of Chemical Engineering}, \orgname{Indian Institute of Technology Bombay}, \city{Mumbai}, \postcode{400076}, \state{Maharashtra}, \country{India}}


\abstract{We demonstrate the use of computer algebra for facilitating the derivation of thin film reduced-order models. We focus on the weighted residual integral boundary layer (WRIBL) method, which has proven to be a very effective technique for developing reduced-order models by averaging the Navier-Stokes equations over the thin-gap direction. In particular, we use SymPy (the symbolic computing library in Python) to derive the core-annular WRIBL model of Dietze and Ruyer-Quil (\textit{J. Fluid Mech. 762, 60, 2015}); the derivation is especially involved due to the inclusion of second-order terms, the presence of two hydrodynamically active phases, the enforcement of interfacial boundary conditions, and the cylindrical geometry. We show, using excerpts of code, how each step of the derivation can be broken down into substeps that are amenable to symbolic computation. To illustrate the application of the derived model, we solve it numerically using scientific computing libraries in Python, and briefly explore the dynamics of the Rayleigh-Plateau instability. The use of open-source computer algebra, in the manner described here, greatly eases the derivation of averaged models, thereby facilitating their use for the study of multiscale flows, as well as for computationally-efficient prediction and optimization.
% by averaging from the Navier-Stokes and continuity equations. }
}




\maketitle

\section{Introduction}\label{sec1}
 The flow in a falling-film evaporator and that in a mucus-lined lung airway share a common aspect: they both belong to the class of thin film flows which is encountered in a variety of industrial, physiological, and geophysical settings \citep{Oron1997long,Matar2009}. Such flows are characterized by a cross-film (transverse) length scale that is much smaller than the streamwise (longitudinal) length scale; the ratio of the two scales is a smaller parameter, $\epsilon \ll 1$, which can be used as a basis for deriving an asymptotic model. In classical lubrication theory, inertia is ignored and the Navier-Stokes equations are truncated at $\mathcal{O}(\epsilon)$, leaving only the longitudinal momentum equation, wherein the transverse viscous diffusion of momentum balances the longitudinal pressure gradient (and other driving forces like gravity). Treating the film height as a parameter, this equation can be solved to yield a self-similar velocity profile; the action of surface tension enters via the application of the stress boundary conditions at the free surface. Enforcing mass balance through the continuity equation then yields a single evolution equation for the height profile of the film. This lubrication equation has been used to study a wide range of problems, including coating flows, Marangoni flows, and dewetting \citep{Oron1997long}. However, lubrication theory fails to describe dynamics that are driven by the film's inertia. The archetypal example is the Kapitza instability and the formation of waves on a gravity-driven falling film \citep{Kalliadasis}.  Indeed, inertia is entirely excluded from the lubrication equation at leading order, and attempts to include inertial effects by extending the single evolution equation to higher orders result in an unphysical model with finite-time blow up \cite{benney1966long,Bala2003,RQcomm2004,Balareply2004,Bala2005}. 
 
The key to the development of inertial thin film models is the addition of a second dynamic field in the evolution equations. Rather than the flow rate being determined by the local film height, one allows the traverse-averaged flow rate to evolve via a dynamic evolution equation of its own (an approximate momentum conservation equation) \citep{RQcomm2004,Balareply2004,Bala2005}, which is of course coupled to the evolution equation for the height (that exactly enforces mass conservation). Different approaches have been used to develop such two-mode models. Two particularly successful techniques are those based on the centre manifold reduction \citep{Roberts96,Roberts-nonnewtonian,Roberts_Li_curved,Robertsbook} and on weighted integration or averaging \citep{ruyer2000improved,Kalliadasis}. Here, we will focus on the latter. Now, naively integrating over the longitudinal momentum equation does yield an equation for the flow rate, but one that contains leading-order errors due to the closure assumption that replaces the true velocity profile by an approximation. An elegant resolution to this difficulty was developed by \citet{ruyer2000improved}: the integration is performed using weight functions that are designed to eliminate the leading-order error and to close the equations at the desired order in $\epsilon$. Since then, various refinements to this weighted-residual integral boundary layer (WRIBL) approach have been developed \citep{Kalliadasis}. The power of this technique is amply demonstrated in its application to two-layer immiscible flows, in stratified \cite{Dietze2013} and core-annular \cite{Dietze2015} configurations. The corresponding WRIBL models have been shown to agree very well with direct numerical simulations based on the volume-of-fluid method \citep{Dietze2013,Dietze2015}. With the aid of such an accurate, physically-consistent, reduced model, one can perform detailed analyses and extensive simulations to gain new physical insights, as exemplified by work on capillary waves \citep{dietze2016kapitza} and secondary instabilities \citep{Dietze2018sliding}, on the effects of rotation \citep{Wray2020-rotation} and electrostatic forcing \citep{Wray2017-electric,Pillai2018,Pillai2020}, and on occlusion \citep{Dietze2020occlusion,Dietze2024plugs,hazra2025probabilistic} and aerosol transport \citep{Swarnaditya-particles} in lung airways. 

The WRIBL procedure can be seen as a general approach to the reduced order modelling of spatially-extended multiscale systems. In fact, it has been applied to thin-gap problems without a free surface, such as the Hele-Shaw flow of fluid with non-negligible inertia \citep{RuyerQuil-heleshaw}, or with a temperature-dependent viscosity that gives rise to thermoviscous fingering \citep{Pillai-thermoviscous}.
 
% The process starts with decomposing the velocity field into two parts: a leading order term which is followed by a correction term. The leading-order velocity is chosen to match the creeping flow profile that is locally fully developed. After substituting the velocity terms into the governing equations, the resulting equations are multiplied by a suitable weight function and are integrated across the film thickness. In addition, the weight function is chosen in such a way that all the correction terms get eliminated from the governing equations or replaced by leading order terms. Finally, we will get a system of coupled partial differential equations. But now the partial differential equations are not enslaved by the liquid film’s thickness rather the system is depend on two variables: the liquid flow rate and the film thickness. 

 The primary challenge in applying the WRIBL method arises from the involved algebraic calculations associated with the derivation; for example, evaluating the weighted residuals requires the analytical calculation of several integrals. The algebra becomes especially unwieldy when the model is extended to $O(\epsilon^2)$ and applied to two-phase flows \citep{Dietze2013,Dietze2015}. The corresponding WRIBL model equations have coefficients that would take multiple pages to write out by hand. The derivation then becomes feasible only with the aid of symbolic computing or computer algebra. Using the computer to perform the algebraic calculations minimizes the chance of errors, while making it easy to derive models for different boundary conditions and driving forces. The goal of this article is to demonstrate---in a pedagogical and step-wise manner---the use of computer algebra in deriving the WRIBL thin-film model. We focus on the particularly challenging case of core-annular two-phase flow; here the derivation is especially involved due to the inclusion of second-order terms, the presence of two hydrodynamically active phases, the enforcement of interfacial boundary conditions, and the cylindrical geometry. 
 % wherein the cylindrical coordinate system adds an additional layer of complexity to the calculations. 
 The steps of our derivation will mirror those in \citep{Dietze2015}. However, each step has to be broken down into simpler substeps, before it can be implemented in a computer algebra system. We explain this procedure in detail, using examples accompanied by excerpts of computer code.

Several computer algebra systems are available, including Mathematica\cite{mathematica}, Maple\citep{maple}, Matlab's Symbolic Math Toolbox \citep{symbolicToolbox}, and Sage-math\cite{sagemath}. Here, we use SymPy\cite{sympy}, the open-source symbolic computing library in Python. Not only is this library freely available (with extensive online documentation), its existence within the rich scientific computing environment of Python allows one to leverage other powerful Python libraries to numerically solve the WRIBL model (with NumPy\citep{harris2020array} and SciPy\citep{SciPy-NMeth}) and to visualize the results (with Matplotlib\citep{Hunter:2007}). The WRIBL equations take the form of partial differential equations with nonconstant coefficients, which after spatial discretization yield highly stiff ordinary differential equations. The advanced time-steppers provided by the \textit{solve{\textunderscore}ivp} function of SciPy greatly help in numerically integrating these equations. Moreover, being open source, these Python libraries can be easily installed and run on clusters whenever parallelization is required for faster computing. 

Note that even though we adopt SymPy for the purpose of illustration, the step-wise procedure we describe here can be implemented in any computer algebra system. So this article should be of general utility to researchers interested in applying the WRIBL approach to obtain reduced-order models.

 % Sympy, the python library for symbolic mathematics\cite{sympy} is a kind of computer algebra software. It can deal with most of the fundamental mathematical operations like integration, differentiation, root finding etc. The open source and comprehensive nature makes it user-friendly. Once the final equations are generated that can be solved in pythons by using numerical packages like scipy and numpy. On top of that, the plotting can be done by matplotlib library. As all the libraries are open source, these can be easily run on clusters whenever parallelization is needed. 
 
 We begin, in Sec.~\ref{gov-equ}, with the governing equations of core-annular flow, simplified for the long-wave limit. We then recall the derivation of the second-order WRIBL model of \citep{Dietze2015} in Sec.~\ref{WRIBL}. The implementation of the derivation using computer algebra, in SymPy, is the subject of Sec.~\ref{symbolic}. 
 % Here we have used various symbolic functionalities Dsolve (To solve ODE's), coeff( To extract the coefficients associated with a particular term of interest) and integrate. In some cases, the integrand was too long to evaluate. So there we had to split the integrand into several parts, integrate them separately and put them back. 
 We then provide examples of the application of the WRIBL model in Sec.~\ref{symbolic}. When the WRIBL model is to be solved numerically, the derivation need not be repeated. Rather, the coefficients of the prederived equations can be read from text files, and then simulated using Python libraries; a Jupyter notebook is provided that illustrates this procedure. We end in Sec~\ref{conclusion} with some concluding remarks.

% \textcolor{red}{change the theme}

\section{Long-wave equations for core-annular flow}\label{gov-equ}
\begin{figure}
\centering
\includegraphics[width=0.6\textwidth]{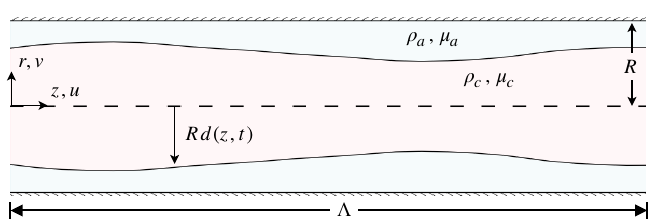}% Here is how to import EPS art
\caption{\label{fig:schematic} Schematic of an axisymmetric core-annular flow in a cylindrical coordinate system.}
\end{figure}

Consider an axisymmetric core-annular arrangement (Fig.~\ref{fig:schematic}) of two immiscible liquids flowing through a cylindrical tube of radius $R$. Both fluids are assumed to be Newtonian, with viscosities \( \mu_c \), \( \mu_a \), and densities \( \rho_c \), \( \rho_a \), where the subscripts \( c \) and \( a \) denote the core and annular phases, respectively. In a cylindrical coordinate system, the inter-fluid interface, which has an interfacial tension \( \gamma \), is located at a distance $R\,d(z,t)$ from the centreline. The length scale of axial variations $\Lambda$ is assumed to be much greater than the radius $R$, so that the ratio  $\epsilon = {R}/{\Lambda} \ll 1$. Scaling the continuity and Navier-Stokes equations, and retaining terms up to \( O(\epsilon^2) \) results in the following long wave equations \citep{Dietze2015} (also called `boundary layer equations' for historical reasons \citep{Kalliadasis}): 

\begin{equation}\label{continuity}
\frac{1}{r}{\partial_r}(r v_i) + {\partial_z u_i} = 0,
\end{equation}
\begin{equation}\label{BLE_main}
 \epsilon Re_i\left(\partial_t u_i + v_i\partial_r u_i+ u_i\partial_z u_i\right)= -\partial_z \left(p_i |_{d}\right) -\epsilon^2 \partial_z \left(\partial_z u_i  |_{d}\right) +\frac{1}{r}\partial_r \left(r\partial_r u_i\right)+2\epsilon^2 \partial_{zz} u_i+\mathcal{B}_i,
\end{equation}
where $Re_i=RU\rho_i/\mu_i$ are the Reynolds numbers and $\mathcal{B}_i = b_iR^2/\mu_i U$ are non-dimensional body forces acting in the $z$-direction in the two phases. The body force could be due to gravity ($b_i = \rho_i g$) or due to an applied pressure gradient. The velocity scale $U$ is problem-dependent: In the presence of an axial driving force, $U$ may be taken to be the mean axial velocity in the unidirectional base state, or, in the presence of an instability, one may choose a velocity scale associated with the fastest-growing mode (which, in case of the Rayleigh-Plateau instability, is proportional to the capillary velocity ${\gamma/\mu_a}$).  The following characteristic scales (denoted by $*$) have been used for non-dimensionalization:
\begin{equation}\label{scales}
  r^* = {R},\, z^* ={\Lambda},\, t^* = {R/U},\,
  d^* = {R},\,  u^*_i = U, \,  v^*_i =\epsilon {U}, \, p_i^* = \frac{\mu_i U}{\epsilon R},
\end{equation}

In Eqs.~\eqref{continuity}-\eqref{BLE_main}, $\partial_t$ denotes the partial derivative with respect to time $t$, with analogous interpretations holding for $\partial_r$ and $\partial_z$. The subscript $i = a, c$ denotes quantities in the annular and core phases. The quantities \( p_i|_d \) and \( u_i|_d \) correspond to the pressure and velocity evaluated at the interface, i.e., at $r = d(z,t)$; these terms arise after substituting for \( \partial_z p_i \) using expressions derived by integrating the \( O(\epsilon^2) \) radial momentum equation from \( r \) to \( d \) \citep{Dietze2015}.

Next, we list all the boundary conditions, starting with those at the interface, across which the velocity must be continuous: 
\begin{equation}\label{vel_con}
u_a = u_c , \quad v_a = v_c, \quad at \quad r = d,
\end{equation}
The tangential and normal stresses must also balance at the interface. Retaining terms to $\mathcal{O}(\epsilon^2)$ yields
 \begin{equation}\label{tangential}
   \partial_r u_a- \Pi_{\mu} \partial_r u_c = \left[2\epsilon^2 \partial_z d\left(\partial_z u_a- \partial_r v_a\right)-\epsilon^2 \partial_z v_a\right] -  \Pi_{\mu}\left[2\epsilon^2 \partial_z d\left(\partial_z u_c-\partial_r v_c\right)-\epsilon^2 \partial_z v_c\right],
 \end{equation}
\begin{equation}\label{normal}
 p_a - \Pi_{\mu} p_c =  - Ca (\kappa) + 2\epsilon^2\left(\partial_r v_a- \partial_r u_a \partial_z d \right)
-2\epsilon^2  \Pi_{\mu}\left(\partial _r v_c- \partial_r u_c \partial_z d \right),
 \end{equation}
 where {$ \Pi_{\mu} = \mu_c/\mu_a$}, $Ca = \gamma/\mu_a U $ is the capillary number, and the mean-curvature $\kappa$ to  $\mathcal{O}(\epsilon^2)$ is given by
\begin{equation}\label{curvature}
  \kappa = \frac{1}{d}-\frac{\epsilon^2\left(\partial_z d\right)^2}{2d}-\epsilon^2 \partial_{zz}.
\end{equation}
This approximation of the full curvature is sufficient to capture the onset of liquid-bridge formation \citep{Dietze2015,hazra2025probabilistic} and the dependence of the shape of stable unduloids (or collars) on the volume of contained liquid \citep{Swarnaditya-particles}. 

The evolution of the interface is governed by the kinematic boundary condition,
\begin{equation}\label{kinematic-gov}
\partial_t d = v_a - u_a\partial_z d.   
\end{equation}

Turning to the centreline, we apply the symmetry condition,
\begin{equation}\label{symmetry}
 v_c = 0, \quad \partial_r u_c = 0 , \quad at \quad r = 0,
\end{equation}
while at the wall, we have the non-penetration  and no-slip conditions,
\begin{equation}\label{no_pen_no_slip}
 v_a = 0, \quad u_a = 0, \quad at \quad r = 1.
\end{equation}
Eqs.~\eqref{continuity}-\eqref{no_pen_no_slip} are a closed system of equations for core-annular flow. Though simplified by discarding terms smaller than $\mathcal{O}(\epsilon^2)$, these equations are still two-dimensional and defined on a deforming domain. The WRIBL method, outlined in the next section, averages across the radial direction and yields one-dimensional equations that depend only on $z$ and $t$.

\section{The WRIBL method of Dietze and Ruyer-Quil \citep{Dietze2015}}\label{WRIBL}
We now recall the WRIBL method which averages the long-wave equations to yield evolution equations for the interface profile $d(z,t)$ and the cross-section averaged flow rates $2\pi Q_i$ of the two phases. 
% {In what follows, we assume that $Re_i \sim \mathcal{O}(\epsilon)$ or smaller and that $Ca \sim \mathcal{O}(1)$.}
{The derivation considers flows for which the Reynolds numbers are not large and so $\mathcal{O}(Re_i\epsilon^2)$ terms are neglected.}

The method begins by decomposing the axial velocity $u_i$ into two parts: a leading term $\hat{u}_i$ and a correction term $u'_i$:
\begin{equation}\label{vel_decomp}
    u_i(r,z,t) = \hat{u}_i(r;d,Q_i) + u'_i(r,z,t),
\end{equation}
The relatively fast viscous diffusion of momentum across the thin film causes the velocity profile to relax quickly to a quasi-steady and quasi-developed profile---the leading order term---with slow time and axial variations that occur only via changes in $d$ and $Q_i$. Since this leading-order velocity profile arises from a balance between transverse momentum diffusion and axial driving forces, we set
% profile arises from balancing the dominant forces in the film, namely the traverse viscous diffusion and the axial pressure gradient: term is a locally steady and developed profile  has an  $\mathcal{O}(1)$ contribution, which is chosen to be locally parabolic, balancing the streamwise pressure gradient with the cross-stream viscous diffusion. Ultimately, the leading-order velocity term contributes solely to the flow rate,  $2\pi Q_i$. 
% To determine the leading-order profile, we must solve the following set of equations..
\begin{equation}\label{u_hat}
  \quad \frac{1}{r}\partial_r \left(r\partial_r \hat{u}_c\right) = A_c, \quad \frac{1}{r}\partial_r \left(r\partial_r \hat {u}_a\right) = A_a,
\end{equation}  
where the $r$-independent forcing terms $A_c$ and $A_a$ are chosen such that the flow rate is determined entirely by $\hat{u}_i$ 
\begin{align}\label{int_cons}
  \quad \int_{0}^{d} \hat{u}_c \,rdr  = Q_c , \quad \int_{d}^{1} \hat {u}_a \,rdr = Q_a.
\end{align}
We also require $\hat{u}_i$ to satisfy the boundary conditions at $\mathcal{O}(1)$, i.e., the boundary conditions obtained on setting $\epsilon \rightarrow 0$:
\begin{align}
   \hat{u}_{a}=0, \quad \hat{v}_{a} = 0\;\;\;  \quad \mathrm{at} \quad r &= 1, \label{uhat_bc_wall}\\
   \
     \partial_r \hat{u}_c = 0, \quad \hat{v}_{c} = 0\;\;\; \quad \mathrm{at} \quad r &= 0, \label{uhat_bc_centreline}\\
    \
        \partial_r \hat {u}_a =  \Pi_{\mu}\partial_r \hat {u}_c, \quad \hat {u}_c = \hat {u}_a  \quad \mathrm{at}   \quad r &= d,\label{uhat_bc_interface}
\end{align}
Solving Eqs.~\eqref{u_hat}-\eqref{uhat_bc_interface} yields the leading contribution to the axial velocity in terms of $Q$ and $d$. 
The velocity corrections $u'_i$, which are atmost of order $\epsilon$, remain undetermined at this stage and, indeed, will not have to be calculated thanks to a judicious choice of weight functions. However, we will make use of the following guage condition that follows from Eq.~\eqref{int_cons}: 
\begin{align}\label{guage}
  \quad \int_{0}^{d} u'_c \,rdr  = 0 , \quad \int_{d}^{1} {u'}_a \,rdr = 0.
\end{align}
Note: The fact that $\hat{u}_i$ is required to exactly yield the flow rate at every point along the domain (see \eqref{int_cons}) implies that the leading-velocity profile has $O(\epsilon)$ adjustments to the purely $O(1)$ profile (which by itself would not be able to account for the $O(\epsilon)$ longitudinal variations of the flow rate). This in turn suggests that alternative choices for $\hat{u}_i$ are possible: one could introduce $O(\epsilon)$ alterations to the leading velocity profile (while making concomitant alterations in the problem for the correction $u'_i$) and thus obtain different weighted-residual models. From this point of view, the choice of $\hat{u}_i$ is a closure assumption, whose suitability must be judged by the performance of the model. Recent work by \citet{Usha2022} has shown how effective reduced-order models can be derived by treating $\hat{u}_i$ as an adjustable profile, whose form is determined by requiring the model to reproduce key asymptotic results of the exact equations. 

Next, we consider the radial velocity, which is also decomposed as
\begin{equation}\label{vel_decomp_radial}
    v_i(r,z,t) = \hat{v}_i(r;d,Q_i) + v'_i(r,z,t),
\end{equation}
where again the corrections $v'_i$ are at most of order $\epsilon$. In the derivation, we shall only require the leading contributions $\hat{v}_i$, which are obtained by integrating the leading-order continuity equation \eqref{continuity}:
\begin{align}\label{v-comp-velocity}
  \quad \hat {v}_c = -\frac{1}{r}\int_{0}^{r} \partial_z\hat{u}_c rdr ,\quad \hat {v}_a = \frac{1}{r}\int_{r}^{1} \partial_z\hat{u}_a rdr
\end{align}
The requirement that $\hat{v}_i$ satisfy the leading-order continuity of velocity condition at the interface,
\begin{equation}
    \hat{v}_a = \hat{v}_c \quad \mathrm{at}   \quad r = d,\label{vhat_bc_interface}
\end{equation}
leads to the following condition on the flow rates $Q_i$:
\begin{equation}\label{flowbalance}
    \partial_z Q_a + \partial_z Q_c = 0
\end{equation}
This is clearly a mass conservation equation, which also follows on multiplying the continuity equation \eqref{continuity} by $r$, integrating across the radial direction, and using Eq.~\eqref{vel_con} and Eq.~\eqref{int_cons}.

We are now ready to average across the momentum equations \eqref{BLE_main}. As the name of the method suggests, we will perform the average using a weight function in each phase, chosen so as to exactly close the dominant viscous diffusion term $r^{-1}\partial_r(r\partial_ru_i)$. However, many unclosed terms will remain, such as those associated with inertia and longitudinal viscous diffusion. This is where the velocity decompositions of Eq.~\eqref{vel_decomp} and Eq.~\eqref{vel_decomp_radial} will play a key role. We shall find that the unclosed terms are all of order $\epsilon^2$, and so on applying Eq.~\eqref{vel_decomp} and Eq.~\eqref{vel_decomp_radial} we will be left with unclosed terms involving $u_i'$ and $v_i'$ that are of order $\epsilon^3$ and so can be neglected in an $\mathcal{O}(\epsilon^2)$ averaged model. In anticipation of this outcome, we first replace $u_i$ and $v_i$ in Eq.~\eqref{BLE_main} by their decompositions in Eq.~\eqref{vel_decomp} and Eq.~\eqref{vel_decomp_radial} and retain only terms up to $\mathcal{O}(\epsilon^2)$:
\begin{multline}\label{BLE}
 \epsilon Re_i\left(\partial_t \hat u_i + \hat v_i\partial_r \hat u_i+ \hat u_i\partial_z \hat u_i\right)=\\ \frac{1}{r}\partial_r \left(r\partial_r \hat u_i\right) + \frac{1}{r}\partial_r \left(r\partial_r u'_i\right) + 2\epsilon^2 \partial_{zz} \hat u_i -\epsilon^2 \partial_z \left(\partial_z \hat u_i  |_{d}\right)
 -\partial_z \left(p_i |_{d}\right)+\mathcal{B}_i
\end{multline}
Recall that $Re_i \sim \mathcal{O}(\epsilon)$ or smaller, as are $u_i'$ and $v_i'$. We perform the same substitution in the boundary conditions. After using the conditions on $\hat{u}_i$ and $\hat{v}_i$ in Eqs.~\eqref{uhat_bc_wall}-\eqref{uhat_bc_interface} and Eq.\eqref{vhat_bc_interface}, we obtain, at the interface ($r=d(z,t)$), 
\begin{gather}
 p_a - \Pi_{\mu} p_c =  - Ca (\kappa) + 2\epsilon^2\left(\partial_r \hat v_a 
- \Pi_{\mu}\partial _r \hat v_c \right), \label{normal_hat}\\
   \partial_r u'_a- \Pi_{\mu} \partial_r u'_c = \left[2\epsilon^2 \partial_z d\left(\partial_z \hat{u}_a- \partial_r \hat v_a\right)-\epsilon^2 \partial_z \hat v_a\right] -  \Pi_{\mu}\left[2\epsilon^2 \partial_z d\left(\partial_z \hat{u}_c-\partial_r \hat v_c\right)-\epsilon^2 \partial_z \hat v_c\right], \label{tangential_prime}\\
   u'_a = u'_c , \quad v'_a = v'_c, \label{vel_cont_prime}
 \end{gather}
and at the centreline ($r=0$),
\begin{equation}\label{symmetry_prime}
 v'_c = 0, \quad \partial_r u'_c = 0,
\end{equation}
and at the wall ($r=1$),
\begin{equation}\label{no_pen_no_slip_prime}
 v'_a = 0, \quad u'_a = 0.
\end{equation}
 
To average across Eq.\eqref{BLE}, denoted henceforth as $BLE_i$, we evaluate the residual $\langle BLE|w \rangle$, where the inner product is defined as $\langle p|q \rangle =  \Pi_{\mu} \int_0^d{p_c q_c  r\,dr} + \int_d^1{p_a q_a r\,dr}$ and $w_i$ are weight functions (yet to be specified). We obtain
\begin{multline}\label{BLE_innerprod}
 \Pi_{\mu}\int_0^d \epsilon Re_c\mathcal{I}_c(\hat{u}_c,\hat{v}_c)w_crdr+\int_d^1 \epsilon Re_a\mathcal{I}_a(\hat{u}_a,\hat{v}_a)w_ardr
  = \\ -\Pi_{\mu}\epsilon^2 \partial_z \left(\partial_z \hat u_c  |_{d}\right)\int_0^dw_crdr-\epsilon^2 \partial_z \left(\partial_z \hat u_a  |_{d}\right)\int_d^1 w_ardr+2\Pi_{\mu}\epsilon^2 \int_0^d \partial_{zz} \hat u_cw_crdr+2\epsilon^2 \int_d^1\partial_{zz} \hat u_aw_ardr  \\+\Pi_{\mu}\int_0^d\partial_r \left(r\partial_r \hat u_c\right)w_cdr +\int_d^1\partial_r \left(r\partial_r \hat u_a\right)w_adr 
 + \Pi_{\mu}\int_0^d\partial_r \left(r\partial_r u'_c\right)w_cdr +\int_d^1\partial_r \left(r\partial_r u'_a\right)w_adr \\
 -\Pi_{\mu}\partial_z \left(p_c |_{d}\right)\int_0^dw_crdr-\partial_z \left(p_a |_{d}\right)\int_d^1w_ardr+\Pi_{\mu}\mathcal{B}_c\int_0^dw_crdr+\mathcal{B}_a\int_d^1w_ardr,
\end{multline}
where $\mathcal{I}_i(\hat{u}_i,\hat{v}_i) = \partial_t  + \hat v_i\partial_r + \hat u_i\partial_z$. The velocity correction $u'_i$ appears in just two terms, which on being integrated-by-parts are recast as follows
\begin{multline}\label{prime_by_parts1}
    \Pi_{\mu}\int_0^d\partial_r \left(r\partial_r u'_c\right)w_cdr +\int_d^1\partial_r \left(r\partial_r u'_a\right)w_adr =\\ \left(-r w_a|_d\partial_ru'_a|_d+\Pi_{\mu}rw_c|_d\partial_ru'_c|_d\right)+ \left(-r\partial_rw_au'_a|_1+\Pi_{\mu}r\partial_rw_cu'_c|_0\right)+\left(r\partial_rw_au'_a|_d-\Pi_{\mu}r\partial_rw_cu'_c|_d\right)\\+(r\partial_ru_a'w_a|_1-\Pi_{\mu}w_cr\partial_r u'_c|_0)+\Pi_{\mu}\int_0^d\partial_r \left(r\partial_r w_c\right)u'_cdr +\int_d^1\partial_r \left(r\partial_r w_a\right)u'_adr
\end{multline}
Keeping in mind the boundary conditions in Eqs.~\eqref{tangential_prime}-\eqref{no_pen_no_slip_prime}, we will find that $u'_i$ can be eliminated from the residual provided the weight functions satisfy the following domain equations and boundary conditions.
% \begin{align}
%    \frac{1}{r}\partial_r \left(r\partial_r w_c\right) = C_c,\quad  &\frac{1}{r}\partial_r \left(r\partial_r w_a\right) = -1\\
%    {w}_{a} = 0\;\;\;  \quad &at \quad r = 1, \\
%    \label{weight_bc_wall}\
%      \partial_r {w}_c = 0 \;\;\; \quad &at \quad r = 0, \\
%     \label{weight_bc_centreline}\
%         \partial_r {w}_a =  \Pi_{\mu}\partial_r {w}_c \quad &at   \quad r = d.
% \end{align}
\begin{align}
   \frac{1}{r}\partial_r \left(r\partial_r w_c\right) = C_c,\quad  &\frac{1}{r}\partial_r \left(r\partial_r w_a\right) = -1 \label{weight_domain}\\
   {w}_{a} = 0\;\;  \mathrm{at} \;\;r = 1, \quad\
     \partial_r {w}_c = 0 \;\; \mathrm{at} \;\; r &= 0,\quad
        \partial_r {w}_a =  \Pi_{\mu}\partial_r {w}_c \;\; \mathrm{at}   \;\; r = d. \label{weight_bc}
\end{align}
The weight functions are not fully specified yet, as $C_c$ is still a free parameter which will be selected shortly. Using integration by parts again, we can simplify the $\mathcal{O}(1)$ integrals of $\hat{u}_i$ in \eqref{BLE_innerprod} as follows:
\begin{equation}\label{hat_by_parts}
    \Pi_{\mu}\int_0^d\partial_r \left(r\partial_r \hat u_c\right)w_cdr +\int_d^1\partial_r \left(r\partial_r \hat u_a\right)w_adr = \Pi_{\mu}C_cQ_c -Q_a
\end{equation}
thereby closing the $\mathcal{O}(1)$ terms exactly. Here we have used the definition of $w_i$ in \eqref{weight_domain}-\eqref{weight_bc} and the boundary conditions on $\hat{u}_i$ in \eqref{uhat_bc_wall}-\eqref{uhat_bc_interface}. After eliminating $u'_i$ using \eqref{prime_by_parts1} along with \eqref{guage} and \eqref{tangential_prime}-\eqref{no_pen_no_slip_prime}, and applying \eqref{hat_by_parts}, we find that \eqref{BLE_innerprod} simplifies to
% Substituting Eqs.~\eqref{weight_domain}, \eqref{weight_bc}, , and \eqref{hat_by_parts} in  \eqref{BLE_innerprod}, and using Eqs.~\eqref{tangential_prime}-\eqref{no_pen_no_slip_prime}, yields
\begin{multline}\label{wribl_C}
 \Pi_{\mu}\int_0^d \epsilon Re_c\mathcal{I}_c(\hat{u}_c,\hat{v}_c)w_crdr+\int_d^1 \epsilon Re_a\mathcal{I}_a(\hat{u}_a,\hat{v}_a)w_ardr
  = \\ -\Pi_{\mu}\epsilon^2 \partial_z \left(\partial_z \hat u_c  |_{d}\right)\int_0^dw_crdr-\epsilon^2 \partial_z \left(\partial_z \hat u_a  |_{d}\right)\int_d^1w_ardr+2\Pi_{\mu}\epsilon^2 \int_0^d \partial_{zz} \hat u_cw_crdr+2\epsilon^2 \int_d^1\partial_{zz} \hat u_aw_ardr  \\-
 dw_a|_d\left[2\epsilon^2 \partial_z d\left(\partial_z \hat{u}_a- \partial_r \hat v_a\right)-\epsilon^2 \partial_z \hat v_a\right] +dw_a|_d \Pi_{\mu}\left[2\epsilon^2 \partial_z d\left(\partial_z \hat{u}_c-\partial_r \hat v_c\right)-\epsilon^2 \partial_z \hat v_c\right]\\+\Pi_{\mu}C_cQ_c -Q_a-\Pi_{\mu}\partial_z \left(p_c |_{d}\right)\int_0^dw_crdr-\partial_z \left(p_a |_{d}\right)\int_d^1w_ardr+\mathcal{B}_a\int_d^1w_ardr+\Pi_{\mu}\mathcal{B}_c\int_0^dw_crdr
\end{multline}
Clearly, $u'_i$ has been eliminated and we are left with an averaged momentum equation involving $d$, $Q_i$, $p_i$ and their derivatives. To obtain an evolution equation for the flow rates, we choose $C_c$ such that 
\begin{equation}\label{wf_int_conts1}
  \int_{0}^{d} w_c rdr  = -\int_{d}^{1} w_a rdr,
\end{equation}
which allows us to replace $\partial_z\left[(p_a - \Pi_{\mu} p_c)|_d\right]$ in \eqref{wribl_C} using the normal stress condition \eqref{normal_hat}. The following equation obtains:
\begin{multline}\label{wribl_noP}
    \Pi_{\mu}\int_0^d \epsilon Re_c\mathcal{I}_c(\hat{u}_c,\hat{v}_c)w_crdr+\int_d^1 \epsilon Re_a\mathcal{I}_a(\hat{u}_a,\hat{v}_a)w_ardr
  = \\ -\Pi_{\mu}\epsilon^2 \partial_z \left(\partial_z \hat u_c  |_{d}\right)\int_0^dw_crdr-\epsilon^2 \partial_z \left(\partial_z \hat u_a  |_{d}\right)\int_d^1 w_ardr+2\Pi_{\mu}\epsilon^2 \int_0^d \partial_{zz} \hat u_cw_crdr+2\epsilon^2 \int_d^1\partial_{zz} \hat u_aw_ardr  \\-
 dw_a|_d\left[2\epsilon^2 \partial_z d\left(\partial_z \hat{u}_a- \partial_r \hat v_a\right)-\epsilon^2 \partial_z \hat v_a\right] +dw_a|_d \Pi_{\mu}\left[2\epsilon^2 \partial_z d\left(\partial_z \hat{u}_c-\partial_r \hat v_c\right)-\epsilon^2 \partial_z \hat v_c\right]\\+\Pi_{\mu}C_cQ_c -Q_a-\left[Ca(\partial_z{k})-2\epsilon^2\partial_z(\partial_r\hat{v}_a-\Pi_{\mu}\partial_r\hat{v}_c)\right]\int_0^dw_crdr+\left(\Pi_{\mu}\mathcal{B}_c-\mathcal{B}_a\right)\int_0^dw_crdr
\end{multline}
After evaluating the integrals and grouping terms based on their $d$ and $Q_i$ dependencies, we arrive at the following evolution equation for $Q_i$:
% The eq\ref{prime_by_parts1} becomes 
% \begin{multline}\label{prime_by_parts2}
%     \Pi_{\mu}\int_0^d\frac{1}{r}\partial_r \left(r\partial_r u'_c\right)w_crdr +\int_d^1\frac{1}{r}\partial_r \left(r\partial_r u'_a\right)w_ardr = w_a|_d\left(-r \partial_ru'_a|_d+\Pi_{\mu}r\partial_ru'_c|_d\right)\\+\Pi_{\mu}\int_0^d\frac{1}{r}\partial_r \left(r\partial_r w_c\right)u'_crdr +\int_d^1\frac{1}{r}\partial_r \left(r\partial_r w_a\right)u'_ardr
% \end{multline}
%and with use of \ref{tangential_prime} finally we have 
% \begin{multline}\label{self_adj}
%     \Pi_{\mu}\int_0^d\frac{1}{r}\partial_r \left(r\partial_r u'_c\right)w_crdr +\int_d^1\frac{1}{r}\partial_r \left(r\partial_r u'_a\right)w_ardr \\= -dw_a|_d\left[2\epsilon^2 \partial_z d\left(\partial_z \hat{u}_a- \partial_r \hat v_a\right)-\epsilon^2 \partial_z \hat v_a\right] +dw_a|_d \Pi_{\mu}\left[2\epsilon^2 \partial_z d\left(\partial_z \hat{u}_c-\partial_r \hat v_c\right)-\epsilon^2 \partial_z \hat v_c\right]\\+\Pi_{\mu}\int_0^d\frac{1}{r}\partial_r \left(r\partial_r w_c\right)u'_crdr +\int_d^1\frac{1}{r}\partial_r \left(r\partial_r w_a\right)u'_ardr
% \end{multline}
% So, 
% \begin{multline}\label{self_adj_hat}
%     \Pi_{\mu}\int_0^d\frac{1}{r}\partial_r \left(r\partial_r \hat{u}_c\right)w_crdr +\int_d^1\frac{1}{r}\partial_r \left(r\partial_r \hat{u}_a\right)w_ardr =\Pi_{\mu}\int_0^d\frac{1}{r}\partial_r \left(r\partial_r w_c\right)\hat{u}_crdr +\int_d^1\frac{1}{r}\partial_r \left(r\partial_r w_a\right)\hat{u}_ardr \\=\Pi_{\mu}C_cQ_c -Q_a
% \end{multline}
\begin{multline}\label{wribl_q_eq}
 \Pi_{\mu i}Re_i\biggl(S_{ij}\partial_t Q_j+F_{ijk} Q_j{\partial_z Q_k}+G_{ijk}Q_jQ_k\partial_z{d}\biggr)\\
 =\Pi_{\mu}C_cQ_c-Q_a-Ca\left(\partial_z\kappa\right)I+\left[\Pi_{\mu}\mathcal{B}_c-\mathcal{B}_a\right]I+J_{j}Q_{j}(\partial_z {d})^2\\+K_{j}\partial_z Q_{j}\partial_z {d}
 +L_{j}Q_{j}\partial_z^2 {d}+M_j\partial_z^2Q_j,
\end{multline}
where summation over repeating indices is implied, and $\Pi_{\mu a} = 1$ and $\Pi_{\mu c} =  \Pi_{\mu}$. The coefficients $S_{ij}, F_{ijk},$ etc., are functions of $d$ alone (and not its derivatives). Determining these coefficients requires one to analytically calculate $\hat{u}_i$, $\hat{v}_i$, $w_i$, and then evaluate the integrals in \eqref{wribl_noP}. This is where computer algebra is indispensable, and the corresponding symbolic calculations are explained in the next section.  

An exact evolution equation for $d$ is obtained by multiplying the continuity equation \eqref{continuity} by $r$ and integrating across the annular phase, from $d$ to $1$. On using the non-penetration condition \eqref{no_pen_no_slip} and the kinematic condition \eqref{kinematic-gov}, we obtain 
\begin{equation}\label{evol_d}
    d\partial_td = \partial_z Q_a
\end{equation}
Equations \eqref{flowbalance}, \eqref{wribl_q_eq}, and \eqref{evol_d} are a closed system for $d$ and $Q_i$, and constitute the second-order WRIBL model.

In many applications, we would like to compute the pressure after solving for $d$ and $Q_i$. Or we may wish to impose boundary conditions on the pressure, rather than on the flow rates. We therefore require an equation for the pressure, which may be obtained by making an alternate choice for the weight function parameter $C_c$. The new choice, denoted as $\Tilde{C}_c$, is such that the corresponding weight functions $\Tilde{w}_i$ satisfy  Eqs.~\eqref{weight_domain}-\eqref{weight_bc} and 
\begin{align}\label{wf_int_conts2}
  \int_{0}^{d} \Tilde{w}_c rdr  = \int_{d}^{1} \Tilde{w}_a rdr 
\end{align}
instead of \eqref{wf_int_conts1}. On using $\Tilde{w}_i$ instead of $w_i$ in \eqref{wribl_C}, and applying the normal stress condition \eqref{normal_hat}, we obtain the following diagnostics equation for $p_c|_d$:
\begin{multline}\label{wribl_p_eq}
 \Pi_{\mu i}Re_i\biggl(\Tilde{S}_{ij}\partial_t Q_j+\Tilde{F}_{ijk} Q_j\partial_z Q_k
 +\Tilde{G}_{ijk}Q_jQ_k\partial_z{d}\biggr)\\
 =\Pi_{\mu}\Tilde{C}_cQ_c-Q_a {-2\Pi_{\mu}\partial_z p_c |_d\Tilde{I}+Ca\left(\partial_z\kappa\right)\Tilde{I}}+\left[\Pi_{\mu}\mathcal{B}_c+\mathcal{B}_a\right]\Tilde{I}+\Tilde{J}_{j}Q_{j}(\partial_z {d})^2\\+\Tilde{K}_{j}\partial_z Q_{j}\partial_z {d}
 +\Tilde{L}_{j}Q_{j}\partial_z^2 {d}+\Tilde{M}_j\partial_z^2Q_j
\end{multline}
Once $p_c|_d$ is known, $p_a|_d$ may be calculated using the normal stress condition \eqref{normal_hat}. {The entire pressure field is then determined, since the $\mathcal{O}(\epsilon^2)$ radial momentum equation implies that $\partial_r(p_i) = 0$.}

\section{Calculating the WRIBL coefficients using computer algebra}\label{symbolic}

To obtain the coefficients $S_{ij}, F_{ijk},$ etc., of the WRIBL equation \eqref{wribl_q_eq}, we must analytically calculate $\hat{u}_i$, $\hat{v}_i$, $w_i$, and then evaluate the integrals in \eqref{wribl_noP}. We begin with $\hat{u}_i$.

\subsection{The leading-order axial velocity $\hat u_i$}
Equations~\eqref{u_hat}-\eqref{uhat_bc_interface} determine $\hat{u}_i$ in terms of $d$ and $Q_i$. We shall first solve \eqref{u_hat}. The excerpt of Python code given below starts by importing the SymPy library (with the tag name `\texttt{sp}') and defining symbolic variables (using the function \texttt{symbols}, called by the phrase `\texttt{sp.symbols}'), including independent variables like $r$, constants like $\Pi_{\mu}$, and functions like $d(z,t)$ and $Q_i(z,t)$. Defining $d$ to be a function of $z$ and $t$ is necessary for SymPy to be able to calculate derivatives like $\partial_z d$. After the definitions, the code specifies the equations and then uses the function \texttt{dsolve} to obtain their general solution. 

\begin{lstlisting}
## python libraries
import sympy as sp
## defining independent variables
r,t,z = sp.symbols('r t z',real = True)
## defining constants
PI_mu = sp.Symbol('PI_mu',positive = True,constant = True)
## defining variables (not meant to be differentiated)
C11,C12,C21,C22 = sp.symbols('C11 C12 C21 C22')
## defining functions of z and t
d      = sp.Function('d',real=True)(z,t)
Uhat_c = sp.Function('Uhat_c',real=True)
Uhat_a = sp.Function('Uhat_a',real=True)
Q_c    = sp.Function('Q_c',real=True)(z,t)
Q_a    = sp.Function('Q_a',real=True)(z,t)
## Differential equations
eq_uhatc = (1/r)*sp.diff(r*sp.diff(Uhat_c(r),r),r)-Ac 
eq_uhata = (1/r)*sp.diff(r*sp.diff(Uhat_a(r),r),r)-Aa 
## DSolve constructs the solutions
soluhatc = sp.dsolve(eq_uhatc)
soluhata = sp.dsolve(eq_uhata)
\end{lstlisting}
The preceding code yields solutions for $\hat{u}_c$ and  $\hat{u}_a$ in terms of two integration constants [Eq.~\eqref{u_hat} is a second order ordinary differential equation]. \texttt{dsolve} names these constants $C_1$ and $ C_2$, by default, in both solutions. To differentiate between the integration constants, we perform the following replacement: $C_{1} \to C_{11}$, $C_{2} \to C_{12}$  for $\hat{u}_c$ and $C_{1} \to C_{21}$, $C_{2} \to C_{22}$ for $\hat{u}_a$. This operation is performed using \texttt{subs}, the substitution function of SymPy. 
%\begin{equation}\label{u_hat_exp}
%  \quad \hat u_c = \frac{A_cr^2}{4}+C_{1}+C_{2}\ln(r), \quad \hat u_a= \frac{A_m r^2}{4}+C_{1}+C_{2}\ln(r),
%\end{equation}
\begin{lstlisting}
## renaming the default integration constants
soluhata = soluhata.subs({(C1,C11),(C2,C12)})
soluhatc = soluhatc.subs({(C1,C21),(C2,C22)})
\end{lstlisting}
%caption={solution for $\hat {u}_a$ and $\hat {u}_c$ }
We will then have the following expressions for $\hat{u}_c$ and $\hat{u}_a$, stored in the symbolic variables `\texttt{soluhatc}' and `\texttt{soluhata}', respectively.
\begin{equation}\label{u_hat_exp}
  \quad \hat u_c = \frac{A_cr^2}{4}+C_{11}+C_{12}\ln(r), \quad \hat u_a= \frac{A_m r^2}{4}+C_{21}+C_{22}\ln(r),
\end{equation}
The next task is to determine the constants in terms of $d$ and $Q_i$ by using the boundary conditions on $\hat{u}_i$ in Eqs.~\eqref{uhat_bc_wall} to \eqref{uhat_bc_interface}. For this, we construct linear equations for the unknowns $C_{11}, C_{12}, C_{21}, C_{22}$ and solve them using the SymPy function \texttt{solve}.
\begin{lstlisting}
## Boundary condition: no slip at the wall (r = 1)
## Extract the expression for \hat{u}_a
uhata1 = soluhata.rhs
## Evaluate the expression at the wall
uhata2 = uhata1.subs(r,1)
##Solve for C11
C11sol = sp.solve(uhata2,C11)
## Replace C11 using the solution obtained above
uhata4 = uhata1.subs(C11,C11sol[0])

## The other constants are obtained in a similar manner
## Boundary condition: symmetry at the core (r = 0) implies that C22 = 0
uhatc1 = soluhatc.rhs
uhatc2 = uhatc1.subs(C22,0) 
## Boundary condition: stress balance at the interface (r = d)
## Differentiate the expression for \hat{u}_a using the function diff()
exp11   = sp.diff(uhata4,r)-PI_mu*sp.diff(uhatc2,r)
exp21   = exp11.subs(r,d) 
C12sol   = sp.solve(exp21,C12)
uhata11 = uhata4.subs(C12,C12sol[0])
## Boundary condition: velocity continuity at the interface (r = d)
exp12   = uhata11-uhatc2 
exp22   = exp12.subs(r,d) 
C21sol   = sp.solve(exp22,C21)
uhatc11 = uhatc2.subs(C21,C21sol[0])

\end{lstlisting}

The solutions for $\hat{u}_i$ thus obtained (stored now in `\texttt{uhata11}' and `\texttt{uhatc11}') contain the unknowns $A_c$ and $A_a$. Our last step then is to calculate $A_c$ and $A_a$ by applying the integral flow rate condition \ref{int_cons}. For this we use the SymPy function \texttt{integrate}.
\begin{lstlisting}
## Flow rate integral constraints
equ1  = sp.integrate(uhata11*r,(r,d,1))-Q_a
equ2  = sp.integrate(uhatc11*r,(r,0,d))-Q_c
solu  = sp.solve((equ1,equ2),(Aa,Ac))
Aasol   = (solu[Aa])
Acsol   = (solu[Ac])
u_c =(uhatc11.subs([(Aa,Aasol),(Ac,Acsol)]))
u_a = (uhata11.subs([(Aa,Aasol),(Ac,Acsol)]))
\end{lstlisting}
We have thus fully determined $\hat{u}_c$ and $\hat{u}_a$ (which are stored in SymPy as `\texttt{u{\textunderscore}c}' and `\texttt{u{\textunderscore}a}'). While the mathematical operations described above are elementary, the algebra would have been tedious to carry out by hand, as evidenced by the lengthy expressions obtained for $\hat{u}_c$ and $\hat{u}_a$; the latter is reproduced below: 
% \begin{multline}\label{uhat_a}
% u\textunderscore{_}a = r^{2} \frac{\left(16 \Pi_{\mu} d^{2}{\left(z,t \right)} \log{\left(d{\left(z,t \right)} \right)} - 4 d^{2}{\left(z,t \right)}\right) Q_{a}{\left(z,t \right)}}{\zeta}\\
% + r^{2}\frac{\left(16 \Pi_{\mu} d^{2}{\left(z,t \right)} \log{\left(d{\left(z,t \right)} \right)} - 8 \Pi_{\mu} d^{2}{\left(z,t \right)} + 8 \Pi_{\mu}\right) Q_{c}{\left(z,t \right)}}{\zeta} \\
% + \frac{\left(- 16 \Pi_{\mu} d^{4}{\left(z,t \right)} + 16 \Pi_{\mu} d^{2}{\left(z,t \right)} + 8 d^{4}{\left(z,t \right)}\right) Q_{a}{\left(z,t \right)}}{\zeta}\log{\left(r \right)} \\
% + \frac{\left(- 8 \Pi_{\mu} d^{4}{\left(z,t \right)} + 16 \Pi_{\mu} d^{2}{\left(z,t \right)} - 8 \Pi_{\mu}\right) Q_{c}{\left(z,t \right)}}{\zeta}\log{\left(r \right)}\\
% +\frac{\left(- 16 \Pi_{\mu} d^{2}{\left(z,t \right)} \log{\left(d{\left(z,t \right)} \right)} + 4 d^{2}{\left(z,t \right)}\right) Q_{a}{\left(z,t \right)}}{\zeta}\\
% +\frac{\left(- 16 \Pi_{\mu} d^{2}{\left(z,t \right)} \log{\left(d{\left(z,t \right)} \right)} + 8 \Pi_{\mu} d^{2}{\left(z,t \right)} - 8 \Pi_{\mu}\right) Q_{c}{\left(z,t \right)}}{\zeta}
% \end{multline}
\begin{multline}\label{uhat_a_solved}
\hat{u}_a = r^{2} 
\frac{
\left(
16 \Pi_{\mu} d^{2} \log\big(d \big) - 4 d^{2} 
\right) Q_a 
}{\zeta} 
+ r^{2} 
\frac{
\left(
16 \Pi_{\mu} d^{2}  \log\big(d \big) 
- 8 \Pi_{\mu} d^{2}  
+ 8 \Pi_{\mu}
\right) Q_c 
}{\zeta} \\
+ \frac{
\left(
- 16 \Pi_{\mu} d^{4}  
+ 16 \Pi_{\mu} d^{2}  
+ 8 d^{4} 
\right) Q_a 
}{\zeta} \log(r)
+ \frac{
\left(
- 8 \Pi_{\mu} d^{4}  
+ 16 \Pi_{\mu} d^{2}  
- 8 \Pi_{\mu}
\right) Q_c 
}{\zeta} \log(r) \\
+ \frac{
\left(
- 16 \Pi_{\mu} d^{2}  \log\big(d \big) 
+ 4 d^{2} 
\right) Q_a 
}{\zeta} 
+ \frac{
\left(
- 16 \Pi_{\mu} d^{2}  \log\big(d \big) 
+ 8 \Pi_{\mu} d^{2}  
- 8 \Pi_{\mu}
\right) Q_c 
}{\zeta},
\end{multline}
where $\zeta$ is given by
\begin{equation*}
    \zeta = \big( 
    4\Pi_{\mu} d ^4 \log(d) 
    - 4\Pi_{\mu} d ^4 
    + 8\Pi_{\mu} d ^2 
    - 4\Pi_{\mu} \log(d ) 
    - 4\Pi_{\mu} 
    - 4 d ^4 \log(d ) 
    + 3 d ^4 
    - 4 d ^2 
    + 1 \big) d ^2.
\end{equation*}
If we focus on the $r$-dependence of $\hat{u}_a$, it can be recast into the following simplified form:
\begin{equation}\label{uhat_a_coeff}
\hat{u}_a = r^2(A_1(d)Q_c+A_2(d)Q_a)
+\log(r)(A_3(d)Q_c+A_4(d)Q_a)+(A_5(d)Q_c+A_6(d)Q_a)
\end{equation}
%The expression for $\hat{u}_a$ is seen to have the form $r^2(A_1(d)Q_c+A_2(d)Q_a)+\dots$,
where the coefficients $A_1, A_2,$ and so on, are functions of $d$ alone. These coefficients will be treated as constants while evaluating integrals with respect to $r$, and so the efficiency of SymPy in performing such integrations can be greatly improved if we represent the solutions of $\hat{u}_i$ (and later of $\hat{v}_i$ and $w_i$) in the form of Eq.~\eqref{uhat_a_coeff}. To do so, we use the \texttt{coeff} function of SymPy. For example, to introduce $A_1$, we first extract the coefficient of $r^2 Q_c$ in \eqref{uhat_a_solved}:
\begin{lstlisting}
## Introducing coefficient variables to simplify expressions
## Extracting coefficient expressions
Ceffr2ua = u_a.coeff(r**2)
Ceffr2uaQc = Ceffr2ua.coeff(Q_c)
\end{lstlisting}
The variable `\texttt{Ceffr2uaQc}' now contains
${\left(16 \Pi_{\mu} d^{2} \log{(d)} - 4 d^{2}\right)}/{\zeta}$. We replace this expression in \eqref{uhat_a_solved} by the function `\texttt{A1}' using the \texttt{subs} function.
\begin{lstlisting}
## replacing coefficient expressions by named functions
A1 = sp.Function('A1', real = True)(d)
u_a = u_a.subs(Ceffr2uaQc,A1)
\end{lstlisting}
We repeat this procedure for all the terms of \eqref{uhat_a_solved} and thereby simplify it to \eqref{uhat_a_coeff}. The same is done for the solution of $\hat{u}_c$ [using coefficient functions named $B_i(d)$] so that it takes on the following form within SymPy:
\begin{equation}\label{uhat_c_coeff}
\hat{u}_c = r^2(B_1(d)Q_c+B_2(d)Q_a)+(B_5(d)Q_c+B_6(d)Q_a)
 \end{equation}
This step, of replacing lengthy expressions of $d$ by named functions, greatly improves the efficiency of SymPy in performing integration and differentiation with respect to $r$. We shall reap the benefits of this strategy now, as we calculate $\hat v_a$ and $\hat v_c$.
%After all such calculations are complete, we will replace the named functions by their full expressions and these operations will be much easier with symbolic functions. Once the operation is over, these coefficients will be substituted back with there actual expression by using sympy commnad "subs". This is going to be demonstrated in the following paragraph during the calculation of $\hat v_a$ and $\hat v_c$.
\subsection{The leading-order radial velocity $\hat v_i$}
To calculate $\hat{v}_i$ using Eq.~\eqref{v-comp-velocity}, we first differentiate $\hat{u}_i$ using the SymPy function \texttt{diff}, to obtain $\partial_z\hat{u}_i$:
\begin{lstlisting}
## symbolic differentiation
duadz= sp.diff(u_a,z)
ducdz= sp.diff(u_c,z)
\end{lstlisting}
Because the coefficient functions like $A_1$ were defined as a function of $d$, SymPy will correctly evaluate the above derivative with respective to $z$ and replace $A_1(d)$ by $\dot{A_1}(d) \partial_z d$ (here the dot denotes the derivative). Next, we perform the indefinite integration in \eqref{v-comp-velocity}:
\begin{lstlisting}
## symbolic integration
v_a = (1/r)*sp.integrate(duadz*r,(r,r,1))
v_c = -(1/r)*sp.integrate(ducdz*r,(r,0,r))
\end{lstlisting}
We now have the solutions for $\hat{v}_i$ in terms of $\dot{A_1}(d), \dot{B_1}(d),$ etc. This form is suitable for subsequent calculations. When we later want to recover the full expressions in terms of $d$ and $Q_i$, we will use the \texttt{subs} function followed by the \texttt{doit} function, to instruct SymPy to evaluate the derivatives while replacing the coefficient functions $A_1(d), B_1(d),$ etc. For example, to replace $A_1$ and $B_1$ in $\hat{v}_a$, one should execute the following: 
\begin{lstlisting}
V_a_temp = sp.expand((v_a.subs([(A1,Ceffr2uaQc),(B1,Ceffr2ucQc).....])))
V_a_full = (V_a_temp).doit()
\end{lstlisting}

% Once $d$ and $Q_i$ are obtained by solving the WRIBL equations, given in the next sub-section, then $\hat{u}_i$ can be calculated to obtain the leading axial velocity profile at any position in the airway. Furthermore, the leading radial-velocity profile can be obtained by integrating the continuity equation \eqref{continuity}:
% \begin{align}\label{v-comp-velocity}
%   \quad \hat {v}_a = -\frac{1}{r}\int_{0}^{r} \partial_z\hat{u}_a rdr ,\quad \hat {v}_m = \frac{1}{r}\int_{r}^{1} \partial_z\hat{u}_m rdr
% \end{align} 
% In this manner, the leading-order, incompressible velocity field $(\hat{u}_i,\hat{v}_i)$ is obtained. \
% To implement this in python

\subsection{The weight function $w_i$}
The weight function $w_i$ is calculated in a similar manner to $\hat{u}_i$. We first solve the second order boundary value problem \eqref{weight_domain}, then use the boundary conditions \eqref{weight_bc} to evaluate the integration constants, and finally calculate $C_c$ using the integral constraint \eqref{wf_int_conts1}. 
%\begin{lstlisting}
%w_a = sp.Function('w_a',real=True)
%w_c = sp.Function('w_c',real=True)  
%eq_wc = (1/r)*sp.diff(r*sp.diff(w_c(r),r),r)-C_c
%eq_wa = (1/r)*sp.diff(r*sp.diff(w_a(r),r),r)-1
%
%## Boundary condition: applying no slip at the wall (r = 1)
%solwa =sp.dsolve(eq_wa)
%wa1 = solwa.rhs
%wa2 = wa1.subs(r,1)
%wa3 = sp.solve(wa2,C1)
%wa4 = wa1.subs(C1,wa3[0])
%
%## Boundary condition: at core (r = 0)
%solwc =sp.dsolve(eq_wc)
%wc1 = solwc.rhs
%wc2 = sp.diff(wc1,r)
%wc3 = wc1.subs(C2,0)
%
%## Boundary condition: at the interface (r = d)
%exp11 = sp.diff(wa4,r)-PI_mu*sp.diff(wc3,r)
%exp21 = exp11.subs(r,d) 
%exp31 = sp.solve(exp21,C2)
%wa11 = wa4.subs(C2,exp31[0])
%
%## Boundary condition: at the interface (r = d)
%exp12 = wa11-wc3
%exp22=exp12.subs(r,d) 
%exp32= sp.solve(exp22,C1)
%wc11 = wc3.subs(C1,exp32[0])
%
%\end{lstlisting}
%Finally the $C_c$ is solved by the \ref{wf_int_conts1} 
%\begin{lstlisting}
%## weight function callculation
%eqwa = sp.integrate(wa11*r,(r,d,1))
%eqwc = sp.integrate(wc11*r,(r,0,d))
%eq   = eqwa+eqwc
%solw = sp.solve(equ,csta)
%wa = wa11.subs(csta,solw[0])
%wm = wm11.subs(csta,solw[0])
%\end{lstlisting}
The result for $w_a$ is given below:
\begin{equation}
   w_a =  \frac{r^{2}}{4} + \frac{1}{2}{\left(\frac{\Pi_{\mu} \left(- d^{4}  + 2 d^{2}  - 1\right)}{\left( 2 \Pi_{\mu} (1-d^{2}) + d^{2} \right) d^{2} } - 1\right) d^{2}  \log{\left(r \right)}} - \frac{1}{4}
\end{equation}
Once again we introduce coefficient functions, $D_1(d), E_1(d)$, etc., to simplify the dependence on $d$ and thus obtain the following representation for $w_i$ in SymPy (stored as `\texttt{w{\textunderscore}a}' and `\texttt{w{\textunderscore}c}'): 
%\begin{lstlisting}
%## coefficient extraction
%Ceffr2wa = w_a.coeff(r**2)
%D1 = sp.Function('D1', real = True)(d)
%w_a = w_a.subs(Ceffr2wa,D1)
%\end{lstlisting}
%Nevertheless the final symbolic expression of weight function for respective phases look like
\begin{equation}\label{w_coeff}
{w}_a = D_1(d)r^2+D_2(d)\ln(r)+D_3, \quad {w}_c = E_1(d)r^2+E_3
 \end{equation}
 
\subsection{The coefficients of the WRIBL model}
With $\hat{u}_i$, $\hat{v}_i$, and $w_i$ in hand, we proceed to calculate the coefficients of the WRIBL model, starting with the inertial terms. For convenience, we recall and expand the left hand side of \eqref{wribl_noP} below.
%\eqref{wribl_q_eq}, i.e., $S_{ij}$, $F_{ijk}$, and $G_{ijk}$. 
% \begin{multline}\label{Inertial part}
%  \Pi_{\mu}\int_0^d \epsilon Re_c\mathcal{I}_c(u_c)w_crdr+\int_d^1 \epsilon Re_a\mathcal{I}_a(u_a)w_ardr = \underbrace{\Pi_{\mu}\epsilon Re_c\int_0^d \partial_t{u_c}w_crdr+\epsilon Re_a\int_d^1 \partial_t{u_a}w_ardr}_\text{term 1}\\+
%  \underbrace{\Pi_{\mu}\epsilon Re_c\int_0^d \hat{u}_c\partial_z{u_c}w_crdr+\epsilon Re_a\int_d^1 \hat{u}_a\partial_z{u_a}w_ardr}_\text{term 2}\\+ 
% \underbrace {\Pi_{\mu}\epsilon Re_c\int_0^d \hat{v}_c\partial_r{u_c}w_crdr+\epsilon Re_a\int_d^1 \hat{v}_a\partial_r{u_a}w_ardr}_\text{term 3}
% \end{multline}
\begin{multline}\label{Inertial-part}
 \Pi_{\mu}\int_0^d \epsilon Re_c\mathcal{I}_c(\hat{u}_c,\hat{v}_c)w_crdr+\int_d^1 \epsilon Re_a\mathcal{I}_a(\hat{u}_a,\hat{v}_a)w_ardr \\= \Pi_{\mu}\epsilon Re_c\left(\underbrace{\int_0^d \partial_t{\hat{u}_c}w_crdr}_{\mathbb{N}_1}+\underbrace{\int_0^d \hat{u}_c\partial_z{\hat{u}_c}w_crdr}_{\mathbb{N}_2}+\underbrace{\int_0^d \hat{v}_c\partial_r{\hat{u}_c}w_crdr}_{\mathbb{N}_3}\right)\\+
\epsilon Re_a \left(\int_d^1 \partial_t{\hat{u}_a}w_ardr+\int_d^1 \hat{u}_a\partial_z{\hat{u}_a}w_ardr+\underbrace{\int_d^1 \hat{v}_a\partial_r{\hat{u}_a}w_ardr}_{\mathbb{N}_4}\right)
\end{multline}
% After performing the integration the above equation gives coefficient $S \,,F\, ,G $.In the following steps we look into these integration, asses them one by one. Finally arrange them in forms of coefficient $S \,,F\, ,G $ 
% % \subsubsection{LHS:}
% \subsubsection{Term 1:}
% \begin{equation}
%     \Pi_{\mu}\epsilon Re_c\left(\int_0^d \partial_t{\hat{u}_c}w_crdr+\int_0^d \hat{u}_c\partial_z{\hat{u}_c}w_crdr+\int_0^d \hat{v}_c\partial_r{\hat{u}_c}w_crdr\right)\nonumber
% \end{equation}
To calculate the integral $\mathbb{N}_1$, we first differentiate $\hat{u}_c$ with respect to time. On examining Eq.~\eqref{uhat_c_coeff}, we see that this operation will yield terms involving $\partial_t Q_i$ as well as $\partial_t d$. Since we wish \eqref{wribl_noP} to yield an evolution equation for $Q_i$, we use the evolution equation for $d$ to replace $\partial_t d$ by $-d^{-1}\partial_z Q_c$ [which follows from Eqs. \eqref{flowbalance} and \eqref{evol_d}]. After performing the integral, we obtain the WRIBL coefficients $S_{ca}$ and $S_{cc}$, of Eq.~\eqref{wribl_q_eq}, by extracting the coefficients of the terms $\partial_t Q_a$ and $\partial_t Q_c$, respectively. In the code below, note the use of the \texttt{expand} function which transforms the integrand into a sum of relatively simple terms, thereby facilitating integration by SymPy. The use of \texttt{expand} also aids in extracting coefficients. The final coefficient expressions are transformed into a compact form using \texttt{simplify}.
\begin{lstlisting}
## evaluating the time derivative
duc_dt = sp.diff(u_c,t)
In1_c_fun = sp.expand(duc_dt*w_c*r)
## substituting the time derivative of the interface profile d
In1_c_fun = In1_c_fun.subs(sp.diff(d,t),(1/(d))*sp.diff(-Q_c,z))
In1_c_fun= sp.expand(In1_c_fun)
## integration of integral N1
In1_c = sp.expand(sp.integrate(In1_c_fun,(r,0,d)))
## Obtaining coefficients S_{cj}
CN1cQc_t = sp.simplify(In1_c.coeff(sp.diff(Q_c,t))) 
CN1cQa_t = sp.simplify(In1_c.coeff(sp.diff(Q_a,t))) 
Scc = CN1cQc_t 
Sca = CN1cQa_t 
\end{lstlisting}
After obtaining the coefficients, we store the corresponding expressions in text files. By reading these text files, a separate code in any computing environment can construct the WRIBL equations, on-demand, for performing either analytical calculations (such as a linear stability analysis) or numerical simulations. Before storing the coefficients, we insert the expressions for $A_1(d)$, $B_1(d)$, etc., and then convert the expression into a string. An optional step that we implement before writing the string to the file 'Scc.txt' is to replace `log' by `np.log'. This is done to facilitate numerical computation in Python using the NumPy library (`\texttt{np}' is the tag we will use when importing NumPy). The code for storing $S_{cc}$ is reproduced below; all other coefficients are stored in an analogous manner.
\begin{lstlisting}
## reintroducing expressions for A1, B1, etc.
Scc = Scc.subs([(A1,Ceffr2uaQc),(B1,Ceffr2ucQc).....])
Scc = str((Scc))
Scc = ((Scc.replace("log","np.log"))) ## for NumPy
## storing in a text file
file= open("./Scc.txt","w") 
file.write(Scc)
file.close()
\end{lstlisting}

Next, we evaluate integrands $\mathbb{N}_2$ and $\mathbb{N}_3$, which will contribute along with $\mathbb{N}_1$ to the coefficients $F_{cjk}$ and $G_{cjk}$ of the terms $Q_j \partial_z Q_k$ and $Q_j Q_k \partial_z d$ in Eq.~\eqref{wribl_q_eq}, respectively.
\begin{lstlisting}
## Integrand N_2
duc_dz = sp.diff(u_c,z)
In2_c_fun = sp.expand(u_c*w_c*r*duc_dz)
In2_c = sp.integrate(In2_c_fun,(r,0,d))
In2_c = sp.expand(In2_c)
## Integrand N_3
duc_dr = sp.diff(u_c,r)
In3_c_fun=sp.expand(v_c*duc_dr*w_c*r)
In3_c = sp.integrate(In3_c_fun,(r,0,d))
In3_c = sp.expand(In3_c)
\end{lstlisting}
To obtain $F_{cjk}$ and $G_{cjk}$, we have to combine the contributions from $\mathbb{N}_1$, $\mathbb{N}_2$, and $\mathbb{N}_3$. This is illustrated below for $F_{cac}$.
\begin{lstlisting}
## contribution from N_1 to F_cac 
CN1cQaQc_z = In1_c.coeff(Q_a*sp.diff(Q_c,z))
## contribution from N_2 to F_cac
CN2cQaQc_z = In2_c.coeff(Q_a*sp.diff(Q_c,z))
## contribution from N_3 to F_cac
CN3cQaQc_z = In3_c.coeff(Q_a*sp.diff(Q_c,z))
## adding the contributions and replacing A1, B1, etc.
Fcac = (CN1cQaQc_z+CN2cQaQc_z+CN3cQaQc_z).subs([(A1,Ceffr2uaQc),(B1,Ceffr2ucQc)...])
\end{lstlisting}
% \subsubsection{ Some WRIBL coeffients from term 1}
% Finally all these coefficients are rearranged and grouped in  \ref{wribl_q_eq} manners. Here we show some of the exmaples
% \begin{lstlisting}
% ## coefficient of Qc_t
% ## Scc = CI1cQc_t

% ## gathering all the coefficients of Qc_z
% # Fccc= CI2cQcQc_z+CI3cQcQc_z+CI1cQcQc_z
% Fccc1 = (CI2cQcQc_z+CI3cQcQc_z+CI1cQcQc_z).subs([(A1,Ceffr2uaQa),(B1,Ceffr2ucQc)...])
% Fccc2 = (Fccc1.subs(d,d1)).doit()
% Fccc3 = str((Fccc2))
% Fccc4 = ((Fccc3.replace("log","np.log")))
% Fccc5 = (Fccc4.replace("d1","d"))
% Aa7= open("./Fccc.txt","w")
% Aa7.write(Fccc5)
% Aa7.close()
% ## gathering all the coefficients of QcQcd_z
% # Gaaa = CI2aQa2d_z+CI3aQa2d_z
% Gaaa1 = ((CI2aQa2d_z+CI3aQa2d_z).subs([(A1,Ceffr2uaQa),(B1,Ceffr2ucQc)...])).doit()
% Gaaa2 = Gaaa1.subs(d,d1)
% Gaaa3 = str((Gaaa2))
% Gaaa4 = ((Gaaa3.replace("log","np.log")))
% Gaaa5 = (Gaaa4.replace("d1","d"))
% Aa16= open("./Gaaa.txt","w")
% Aa16.write(Gaaa5)
% Aa16.close()
% \end{lstlisting}
% All the other terms got extracted and arranged in the similar manners.  
So far we have discussed the evaluation of the integrals multiplying $Re_c$ in Eq.~\eqref{Inertial-part}. The three terms multiplying $Re_a$ have the same form and so the same steps apply, except for term $\mathbb{N}_4$. Because the expressions for $\hat{u}_a$ and $w_a$ are lengthier than those for $\hat{u}_a$ and $w_a$, the integral $\mathbb{N}_4$ requires special treatment that was not needed for $\mathbb{N}_3$ (else the integration takes a very long time).
The integrand of $\mathbb{N}_4$ is split into several partial fractions which are integrated individually and then recombined.   
\begin{lstlisting}
dua_dr = sp.diff(u_a,r)
In3_a_fun = sp.expand(v_a*dua_dr*w_a*r)
## splitting
In3_a_array=(sp.Add.make_args(In3_a_fun))
In3_a_result = []
## integrating each part
for i in range(len(In3_a_array)):
    In3_a_result.append(sp.integrate(In3_a_array[i],(r,d,1)))
## recombining
In3_a = In3_a_result[0]
for j in range(1,len(In3_a_array)):
    In3_a = In3_a + In3_a_result[j]
In3_a = sp.expand(In3_a)
\end{lstlisting}

With the preceding discussion of the calculation of the coefficients of the inertial terms, we have introduced all the techniques needed to obtain the remaining coefficients of the WRIBL model. One simply has to evaluate the integrals in \eqref{wribl_noP}, one at a time, and extract from the results the contributions to the coefficients of \eqref{wribl_q_eq}. 
% Recall that the pressure gradient terms in \eqref{wribl_C} have to be replaced by the normal stress condition before the integrations are performed. 
The coefficients of the WRIBL pressure equation \eqref{wribl_p_eq} are obtained in an analogous manner to those of \eqref{wribl_q_eq}.

Note that the weight functions used here differ slightly from those in \citet{Dietze2015}. We have checked that their WRIBL coefficients match those obtained here, after rescaling to account for the difference in weight functions. We have also validated out simulation results with those of \citet{Dietze2015} [see Fig.~\ref{fig:gs}(a) for an example].
% We have also checked that the predictions of our WRIBL equations match those of \citet{Dietze2015}, as illustrated by figure~\ref{fig:validation} in Appendix~\ref{app:validate}. 

\section{Illustrative simulations of the WRIBL model}\label{application}

We now present simulations of the WRIBL model to illustrate its utility for the study of two-phase flow phenomena. Two different core-annular cases are considered; Table~\ref{prop} lists the corresponding fluid pairs and their properties. Note that while no external driving force is imposed in case 1, we have gravity driven flow in case 2. 
% The less common scenario of a driving force in only the core phase is taken up in case 3; imagine flow through a tube that is prelined with a lubricating film. This third scenario illustrates the ability of the WRIBL model to account for different driving forces in the two phases. 

The cylindrical interface in these core-annular systems is susceptible to the Rayleigh-Plateau instability. The second-order WRIBL model is a potent tool for investigating such flows \citep{Dietze2015,Dietze2020occlusion,Dietze2024plugs,hazra2025probabilistic}. Not only does the model account for inertial effects, it also includes longitudinal viscous diffusion (which is relevant near the necks of draining annular films) and nonlinear interfacial curvature (which is necessary for capturing the transition from open to occluded tubes, caused by the formation of liquid-bridges).

In what follows, we use $U = R/\tau$ as the velocity scale, where $\tau$ is the inertialess approximation of the linear-growth timescale of the fastest-growing instability mode and is given by $\tau = 6 ({ d_0 R\mu_a}/{\gamma})\left(\alpha^4(1/\alpha^2 - 1)(1/d_0-1)^2(1/d_0^2-1)\right)^{-1}$. Here, $\alpha = {2\pi R d_0}/{\Lambda}$, where $R d_0$ is the thickness of the initially flat film, and $\Lambda = 2\pi 2^{1/2} Rd_0$ is the approximate wavelength of the fastest-growing Rayleigh-Plateau mode (this inviscid prediction of \citet{Rayleigh1892wavelength} works very well even for viscous films \citep{Dietze2015}.) Our simulations are performed on an axial domain of non-dimensional length $\Lambda/R$, with periodic boundary conditions. We initiate the simulations by perturbing the flat film with a single sinusoid corresponding to the fastest-growing instability mode: $d = d_0 + 10^{-3}\,\mathrm{cos}(2 \pi z R/\Lambda)$.

To facilitate the simulation of the WRIBL model, we first use \eqref{flowbalance} to substitute $Q_a = Q_t(t)-Q_c$ (where the total flow rate $Q_t$ depends only on $t$) in the flowrate evolution equation \eqref{wribl_q_eq}, as well as in the pressure equation \eqref{wribl_p_eq}. We then integrate the latter across the $z$-direction, and use the fact that $\int_0^{\Lambda/R} \partial_zp_i dz = 0$ for periodic boundary conditions, to obtain an ordinary differential equation (ODE) for $Q_t(t)$:
\begin{equation}\label{Qt_ode}
d_t Q_t = \Phi (d,Q_c,Q_t)
\end{equation}
Equations \eqref{evol_d}, \eqref{wribl_q_eq} and \eqref{Qt_ode} form a closed system, of two partial differential equations (PDEs) and one ODE, for $d$, $Q_c$ and $Q_t$.

\begin{figure}
\centering
\includegraphics[width=.9\textwidth]{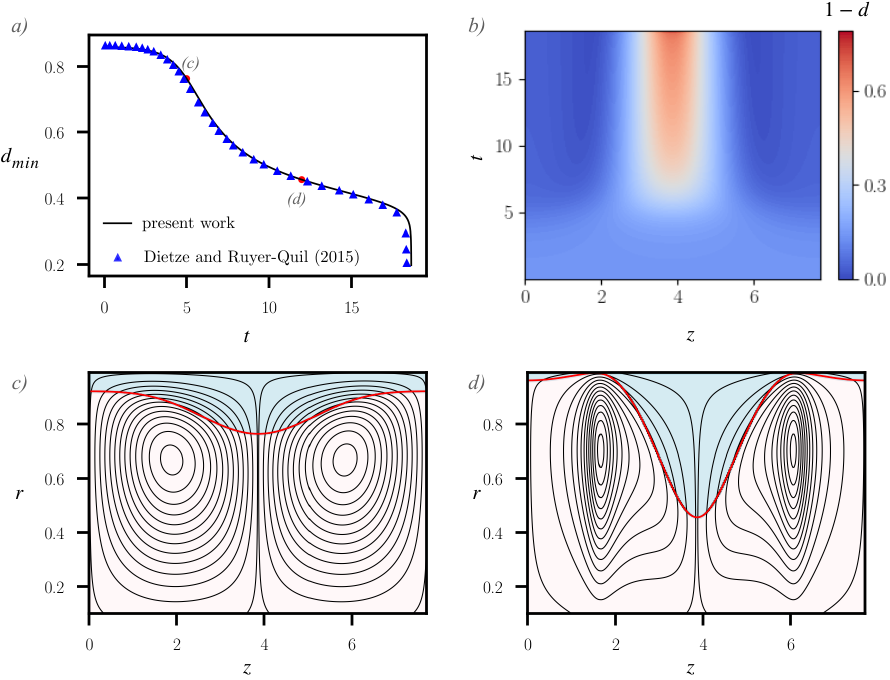}% Here is how to import EPS art
\caption{\label{fig:gs} Rayleigh-Plateau instability of a annular viscous liquid film linning a tube; the fluid properties are chosen to mimic mucus and air in lung airways (case 1 of Table~\ref{prop}). (a) Evolution of the  minimum interface position $d_{min}$. The blue triangles correspond to the results of \citet{Dietze2015}. (b) Space-time kymograph showing contours of the film thickness, $1-d(z,t)$. Panels (c) and (d) present the streamlines in the two phases, in a stationary reference frame, for two time instances [see the labels in panel (a)]. A Jupyter notebook that simulates the WRIBL model and generates this figure is available at \href{https://cocalc.com/share/public_paths/93b59c92b2c7b5b66f21df18b052b5941dff1cd5}{https://cocalc.com/share/public\textunderscore{paths}/93b59c92b2c7b5b66f21df18b052b5941dff1cd5} .}
\end{figure}

We discretize the PDEs in space, using a second-order central-difference scheme---an evenly-spaced spatial grid of 500 points is found to be sufficient for grid-independent solutions. The resulting system of ODEs is integrated in time using the stiff, adaptive time-stepping, LSODA algorithm \citep{hindmarsh1995algorithms}. This method is available as an option in the \texttt{solve{\textunderscore}ivp} function, of the Python library SciPy \citep{scipy}. A link to a Python Jupyter notebook that reads-in the WRIBL coefficients from text files, and then simulates the model using \texttt{solve{\textunderscore}ivp}, is provided in the caption of Fig.~\ref{fig:gs}.  

\begin{table}
\caption{Properties of core-annular fluid systems chosen for illustrating the application of the WRIBL model.}\label{prop}%
\begin{tabular}{@{}clcccccccc@{}}
\toprule
case & core - annular fluids  & $\rho_c/\rho_a$  & $\mu_c/\mu_a$& $\gamma$ ($Nm^{-1}$) & $\mathcal{B}_c$ &$\mathcal{B}_a$ & $Ca$ & $Re_c$ & $Re_a$ \\
\midrule
1 &air - mucus   & 0.001   & 0.0018  & 0.025 & 0 & 0 & 2213.25 & 0.023 & 0.037\\
2 &air - water    & 0.0012   & 0.018  & 0.076 & 5.78 & 86.65  & 7458.82 & 0.203 & 3.05\\
% 3 &silicon oil - water    & 1.19   & 5.6  & 0.035 & 1.14 & 0  & 7458.82 & 1.88 & 8.86\\
%3 &silicon oil - glycerol    & 0.95  & 0.01  & 0.03  \\
\botrule
\end{tabular}
\end{table}

Case 1 (see Table~\ref{prop}) corresponds to a very viscous annular liquid film and a gaseous core; this configuration is an idealization of the mucus linning of pulmonary airways \citep{Levy2014}. For sufficiently high annular fluid volumes, the Rayleigh-Plateau instability causes the annular film to accumulate into a hump that continues to grow until it forms a liquid-bridge or plug, which occludes the airway \citep{everett1972model,johnson1991}. Figure~\ref{fig:gs} illustrates the dynamics that lead up to plug formation. Three regimes of hump-growth, described in \citet{Dietze2015} and \citet{Dietze2018sliding}, are noticeable in the time-trace of the minimum interface position $d_{min}$ in Fig.~\ref{fig:gs}(a): (i) initial growth of the instability, leading to the emergence of a hump [see Fig.~\ref{fig:gs}(c)]; (ii) slowing down due to viscous drag at the necks on either side of the growing hump [see Fig.~\ref{fig:gs}(d)]; (iii) resumption of rapid-growth that ends in plug formation [final rapid decrease of $d_{min}$ in Fig.~\ref{fig:gs}(a)]. Our simulation (solid line) is seen to agree well with that of \citet{Dietze2015} (triangles). 

Figure~\ref{fig:gs}(b) presents the space-time kymograph of the film thickness, $1-d(z,t)$, and shows that the hump does not translate during its evolution. Note, however, that such humps can spontaneously break symmetry and slide due to a secondary instability at the thinning necks of the annular film \citep{Lister2006capillary,Glasner2007,Dietze2018sliding}. 

The streamlines in Figs.~\ref{fig:gs}(a-b) correspond to contours (not equispaced) of the streamfunction $\Psi_i$, which is calculated from the leading-order velocity field:
% \begin{align}\label{streamlines}
% \quad \Psi_a = \int_{r}^{1} \hat{u}_a \,r dr, \quad
% \Psi_c = \int_{0}^{r} \hat{u}_c \,r dr, 
% \end{align}
\begin{align}\label{streamlines}
\quad \Psi_c = \int_{0}^{r} \hat{u}_c \,r dr, \quad
\Psi_a = \Psi_c|_d+\int_{d}^{r} \hat{u}_a \,r dr. 
\end{align}
% provided $\Psi_a|_d = \Psi_m|_d$
The fact that the solution of the WRIBL model, namely the fields $d(z,t)$ and $Q_i(z,t)$, can be used to unambiguously compute the velocity field is one of the advantages of the weighted residual approach \citep{Usha2022}.

What about the dynamics post plug formation? A thin-film model, which represents the interface position as a single-valued function of $z$, cannot be expected to describe the dynamics beyond the coalescence event. It is truly remarkable, therefore, to find that the WRIBL model can describe the formation and motion of liquid plugs, via the use of a disjoining pressure at the centreline (see \citet{Dietze2020occlusion} and \citet{Dietze2024plugs}). In this augmented WRIBL model, the runaway growth of annular humps results in the formation of pseudo-plugs, whose dynamics are in good agreement---both qualitatively and quantitatively---with that of true plugs formed in volume-of-fluid simulations \citep{Dietze2024plugs}. Such quantitative accuracy, given the relatively rapid variations of the interface in the vicinity of pseudo-plugs, is a testament to the efficacy of the WRIBL approach.

\begin{figure}
\centering
\includegraphics[width=.9\textwidth]{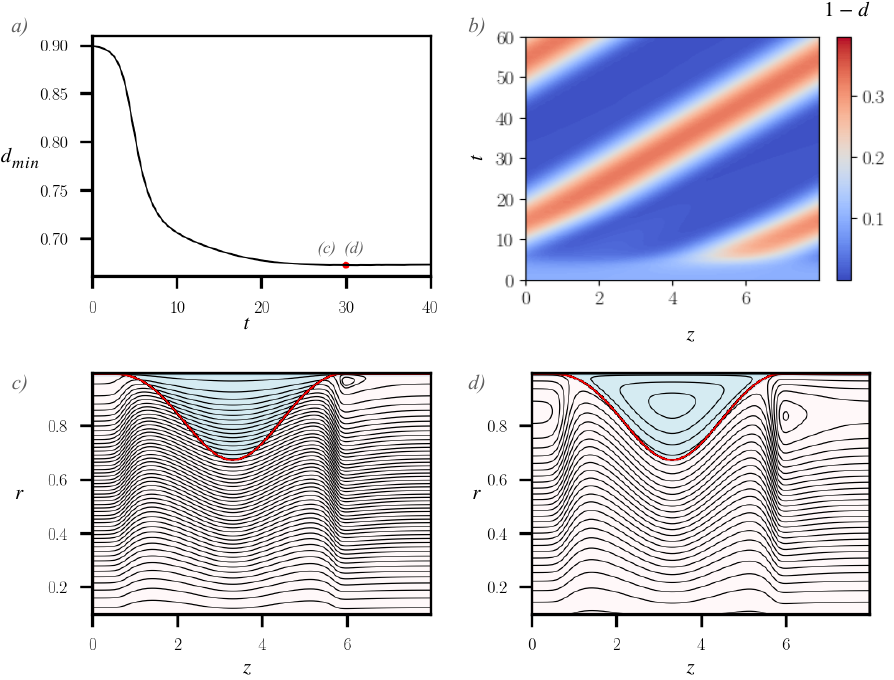}% Here is how to import EPS art
\caption{\label{fig:wso} Gravity driven flow of air (core) and water (annulus) down a vertical tube (case 2 in Table~\ref{prop}). (a) Evolution of the  minimum interface position $d_{min}$. (b) Space-time kymograph showing contours of the film thickness, $1-d(z,t)$. Panels (c) and (d) present the streamlines in the two phases in a stationary frame [panel (c)] and in a frame that moves with the long-time asymptotic speed of the water hump [panel (d)]. The corresponding time instant is indicated by the red marker in panel (a). }
\end{figure}

We now turn to case 2, which corresponds to gravity-driven core-annular flow of air (core) and water (annulus). Gravitational forcing and the associated axial mean-flow promotes the nonlinear saturation of the Rayleigh-Plateau instability \citep{Frenkel87,Quere90}. Moreover, increasing the relative strength of gravitational forcing causes the dynamics to transition from being Rayleigh-Plateau dominated to being governed by the Kapitza instability of falling films \citep{Dietze2020occlusion}; correspondingly, the spatio-temporal character of the
instability changes from absolute to convective. Here, we consider a case of relatively weak gravitational forcing (a narrow tube), such that the Rayleigh-Plateau instability produces a hump, but one that saturates to form a stable annular collar or unduloid \citep{johnson1991,Dietze2015}. This saturation is evident in Fig.~\ref{fig:wso}(a). The unduloid then flows down the tube (like annular drops flowing down a fibre \citep{Quere90}), as seen in the kymograph of Fig.~\ref{fig:wso}(b). The air in the core also flows along with the annular unduloid; the corresponding streamlines are presented in Figs.~\ref{fig:wso}(c). The relative motion in the two fluids is better represented in a moving frame that translates with the long-time asymptotic speed of the unduloid. The corresponding streamlines are shown in Fig.~\ref{fig:wso}(d). We see that liquid circulates within the unduloid; in addition, a vortex is present in the air, near the upstream slope of the unduloid. (A similar gas-phase recirculation zone is present atop the upstream face of Kaptiza waves on a falling liquid film~\citep{Dietze2013}.) 

In the simulation illustrated by Fig.~\ref{fig:wso}, the depleted film outside the unduloid continues to drain into the unduloid, but very slowly. This allows us to continue the simulation for long times and to see the unduloid falling along the tube. However, the simulation would have to be stopped when the interface meets the wall (which will occur in finite time \citep{Lister2006capillary}), because the WRIBL model cannot be simulated once $1-d$ becomes zero. In order to continue the simulation beyond this dryout event, we must include a precusor film at the wall \citep{Oron1997long,Ghatak1999dynamics}. This is typically done by augmenting a thin film model with a disjoining pressure of the form ${h^a_0}{(1-d)^{-b}}$, where $h_0 > 0$, and $a$ and $b$ are positive integers (e.g., $a=6$, $b=9$, and $h_0 \sim 10^{-4}$ yields a precursor film thickness of $\sim 10^{-3}$ \citep{Swarnaditya-particles}). The regions of the wall occupied by the precursor film are then treated as being devoid of liquid.  The precursor film approach has long been used to study the movements of droplets along solid surfaces using thin film models (see \citet{RuyerQuil2023droplets} for a recent example).

Here, we have restricted ourselves to simple initial-value-problem simulations, for the sake of illustrating the application of the WRIBL model. However, one of the key advantages of the reduced-order WRIBL model is that one can perform advanced mathematical and computational analyses which would be much more challenging with the full Navier-Stokes equations. For example, one can use continuation to compute travelling waves \citep{Dietze2013,Dietze2024plugs} and solitons \citep{ruyer2000improved,BalaSoliton2006,Bala2011,Kalliadasis}, perform stability and bifurcation analyses \citep{Wray2020-rotation,Kalliadasis25}, and study the response of the system to periodic forcing \citep{Bala2011,Dietze2013,Pillai2018,Pillai2020,Dietze2020occlusion} and stochastic perturbations \citep{Dietze2020occlusion,hazra2025probabilistic}. 
% By enabling such analyses, reduced models based on the weighted residual approach have deepened our understanding of thin film flows.

% \begin{figure}
% \centering
% \includegraphics[width=.9\textwidth]{image/act_core.pdf}% Here is how to import EPS art
% \caption{\label{fig:core} In this figure we are represent case 2: silicone oil- water as core -annular fluid.  The core is denser and viscous than the annular fluid. The core is also active a) evolution of the interface minima vs time. b) space-time kymograph is depicting the motion of liquid hump under influence of flowing core.  c,d are the streamlines in laboratory reference frame and moving frame respectively at $t(t^*/\tau) = 100$ (marked by red circles in the panel a).}
% \end{figure}

% In our last example (case 3 of table~\ref{prop}), we have a denser and more viscous core fluid. A driving force is applied only to the core fluid, to mimic the flow that would occur when a liquid flows though a tube that is pre-lined with a thin film of lubricant. Our pedagogical reason for including this case is that it demonstrates the ability of the WRIBL model to  how the WRIBL model 

\section{Concluding remarks} \label{conclusion}
The WRIBL averaged model has proven to be very useful for exploring the fluid dynamics of thin films that are strongly influenced by inertia, as well as other physical effects that are neglected in traditional lubrication models (such as longitudinal viscous momentum-diffusion). Simulating the WRIBL model requires far less computation time and resources as compared to direct numerical simulations (DNS) of the full Navier-Stokes equations on a deforming domain. At the same time, the results of the WRIBL model agree very well with DNS \citep{Dietze2013,Dietze2015}. Thus, the WRIBL approach is an exemplar of reduced-order multiscale modelling. In this article, we have attempted to make this modelling approach more accessible by the use of a symbolic computing engine. By performing the involved algebraic calculations on the computer, one can better appreciate the structure of the derivation while reducing the chance for errors. Once the code is setup for deriving a particular version of the WRIBL model, it is then relatively easy to obtain models for different variants of the physical problem, thus enabling exploration of the rich physics of thin film flows. Indeed, the combination of computer-algebra code, for deriving the model, with a numerical-computing library, for solving it, makes it possible to use the WRIBL model in the classroom---especially if the supporting codes are open-source. From a research perspective, an open-source computer algebra code for deriving reduced models makes it easier for other researchers to use the model, and also enables comparison between independent derivations of the model (equality between two expressions, no matter how lengthy, can be verified in the computer algebra system by checking whether their difference simplifies to zero; this is how we validated our WRIBL coefficients with those provided in \citep{Dietze2015}).\\

\backmatter

\bmhead{Supplementary information} 
% Movies accompanying Figs. 2-4 are available online at \href{https://bighome.iitb.ac.in/index.php/s/5EPYPiRXjy3RKsn}{https://bighome.iitb.ac.in/index.php/s/5EPYPiRXjy3RKsn}. 
A Python Jupyter notebook that reads in the WRIBL coefficients from text files and solves the model numerically is provided at \href{https://cocalc.com/share/public_paths/93b59c92b2c7b5b66f21df18b052b5941dff1cd5}{https://cocalc.com/share/public\textunderscore{paths}/93b59c92b2c7b5b66f21df18b052b5941dff1cd5}.

\bmhead{Acknowledgements}
The authors are grateful to Georg Dietze (Lab. FAST, Univ. Paris-Saclay) and Christian Ruyer-Quil (LOCIE, Univ. de Savoie Mont-Blanc) for many helpful discussions. This work was supported by DST-SERB (J.R.P., grant no. SRG/2021/001185) and IRCC, IIT Bombay (S.H., Ph.D. fellowship; J.R.P., grant no. RD/0519-IRCCSH0-021). J.R.P. acknowledges his Associateship with the International Centre for Theoretical Sciences (ICTS), Tata Institute of Fundamental Research, Bangalore,
India. The authors thank the National PARAM Supercomputing Facility \textit{PARAM SIDDHI-AI} at CDAC, Pune for computing resources; simulations were also performed on the IIT Bombay workstation \textit{Gandalf} (procured through DST-SERB grant SRG/2021/001185).
% , and \textit{Faramir} and \textit{Aragorn} (procured through the IIT-B grant RD/0519-IRCCSH0-021).

%%=============================================%%

\FloatBarrier
\bibliography{ref}% common bib file
% if required, the content of .bbl file can be included here once bbl is generated
% \input sn-article.bbl

\end{document}